# A Feedback Model of the Relationship between Cerebral Blood Flow and Arterial Blood Pressure.

Alexander Gersten

Department of Physics and Zlotowski Center of Neuroscience,

Ben-Gurion University of the Negev, Beer-Sheva, Israel

**Address correspondance to**:

Prof. A. Gersten

Dept. of Physics,

Ben-Gurion University of the Negev

P.O.B. 653

84105 Beer-Sheva, Israel

E-mail: gersten@bgumail.bgu.ac.il

Phone: 972-8-6461853

Fax: 972-8-6472904

**Summary:**

The dependence of cerebral blood flow (CBF) on mean arterial blood pressure (MABP) can be mathematically described by a simple model based on the following assumptions: 1) At MABP levels below the lower limit of autoregulation, denoted as $MABP_1$, there are no autoregulatory or feedback mechanisms that influence CBF. Between $MABP_1$ and $MABP_2$, the level of MABP at which breakthrough occurs, there is a linear, MABP-dependent feedback with a slope that is characteristic of the indiviual. . The classical condition of autoregulation with a plateau between $MABP_1$ and $MABP_2$ is a particular case of this model. This model has been found to describe very well well the results of the experiments performed on dogs by Harper et al.(1966) in which the individual feedback slope parameter varied greatly, indicating the importance of measurements in individuals rather than of averages of measurements in different individuals often used to describe autoregulation. The model permits to observe decreased CBF, while increasing MABP, the data of two dogs have this property.

Autoregulation of cerebral blood flow (CBF), i.e., constancy of CBF over a wide range of mean arterial blood pressure (MABP), is generally accepted as a well established fact. Examination of published experimental data, however, fails to provide proper experimental evidence to support the existence of the autoregulatory plateau.

The first publication generally believed to prove the existence of autoregulation in humans was that of Lassen (1959). The data at different points on the plateau were, however, taken from different experimental populations by different groups of investigators, and some elevated MABP was achieved by infusion of drugs that could also constrict the cerebral vessels (Lassen 1959, Reivich 1969). Lassen's analysis may, therefore, serve as an indication but not a proof of cerebral autoregulation.. Subsequently, Harper (1966) studied 12 dogs in each of which he measured CBF over a wide range of MABP. Although he presented and discussed the results taken together on 8 normocapnic and 4 hypercapnic animals ,.he fortunately, did provide the experimental data for each dog separately. These are the only published data that describe individually in each animal the dependence of CBF on MABP over a wide range of MABP. MacKenzie et al. (1979) also measured CBF over a wide range of MABP in baboons, but they provided only averaged data from several baboons in support of autoregulation We shall, therefore, use Harper's data extensively for our analyses that are based on a simple model of autoregulation.

The mechanism of cerebral autoregulation is still not well understood and a subject of controversy. For recent reviews of the experimental and theoretical considerations the reader is encouraged to consult the reviews of Aslid (1989), Gotoh and Tanaka (1988),

Ursino (1991), Sokoloff (1996), and the references therein. The book of Hademnos and Massoud (1998) has complementary material and references.

## METHODS AND MATERIALS

**Generalization of the classical model**

The oversimplified classical picture of autoregulation is depicted in Fig. 1. In this illustration for MABP's in the range of 60-130 mm Hg there is no change in CBF. The range of autoregulation may change somewhat in cases of hypertension and hypotension;. the plateau may be shifted to the right in hypertension and to the left in hypotension. Let us describe Fig. 1 in terms of a simple control feedback model. Without any feedback CBF is expected to rise linearly with MABP

$$CBF(MABP) = S \cdot MABP, \qquad (1)$$

where S is a constant representing the initial linear slope of the CBF with respect to MABP in units of ml/min/100g/mmHg. In order to achieve the plateau, the line of Eq.1 for CBF as a function of MABP must be diminished from the values predicted by Eq.1 by a contribution that is different for each value of , MABP. In a feedback control model this subtracted quantity is equal to $-\Phi(MABP) \cdot CBF(MABP)$, where $\Phi(MABP)$ is a feedback (gain) function that depends on MABP. Eq.1 can now be replaced with

$$CBF(MABP) = S \cdot MABP - \Phi(MABP) \cdot CBF(MABP) . \qquad (2)$$

Transposing,

$$CBF(MABP) + \Phi(MABP) \cdot CBF(MABP) = (1 + \Phi(MABP)) \cdot CBF(MABP) = S \cdot MABP$$

Dividing both sides by $1 + \Phi(MABP)$ yields

$$CBF(MABP) = \frac{S \cdot MABP}{1 + \Phi(MABP)}. \qquad (3)$$

Solving for $\Phi(MABP)$,

$$\Phi(MABP) = \frac{S \cdot MABP}{CBF(MABP)} - 1 \quad . \qquad (4)$$

In many instances the constant S can be relatively well determined from experiments in which CBF is measured at different levels of MABP. In these cases Eq. 4 is very advantageous, as it allows determination of $\Phi(MABP)$ directly from the experiments. We shall extensively employ this procedure in the analyses of Harper's (1966) data.

In order to obtain the continuous rising line and plateau of Fig. 1, the feedback function

$$\Phi(MABP) = \begin{cases} 0, & \text{for} \quad 0 \leq MABP \leq 60mmHg \\ \dfrac{S \cdot MABP}{CBF_{plateau}} - 1, & \text{for} \quad 60mmHg \leq MABP \leq 130mmHg \\ \dfrac{MABP}{MABP - 70mmHg} - 1, & \text{for} \quad MABP \geq 130mmHg \end{cases} \qquad (5)$$

where $CBF_{plateau}$ is the value of CBF at the plateau (here 50 ml/min/100g, for example). It should be stated that the behavior of CBF above 130 mm Hg is uncertain. In the data of MacKenzie et al. (1979) we see a sharp rise in CBF at MABP levels above 130 mm Hg that is not seen in the Harper's data. For reasons of symmetry we have assumed the that the slope of CBF with respect to CBF is the same at levels of MABP above 130 as below 60 mm Hg. That is why we pointed out in Fig. 1 and Fig. the uncertainty of the model with MABP levels above 130 mm Hg. The dependence of $\Phi$ on MABP, as shown by Eq. 5, is illustrated in Fig. 2. It should be noted that in the plateau region its dependence on MABP is linear and thus very simple. We can generalize this model by allowing a linear dependence of $\Phi$ on MABP, but with different strengths. Above MABP of 130 mm Hg one may see in Fig. 2 the breakdown of the linearly growing biofeedback contribution, but it should be recognized that our knowledge of what is going on in this region is quite uncertain.

The classical view of autoregulation is depicted in Fig.1. This is an over-idealized picture, for we know that there are very large deviations from this picture. For example, during hypercapnia there is almost no feedback suppression and, therefore, almost no autoregulation. We shall consider such effects by changing the feedback of Eq.2 by

$$CBF(MABP) = S \cdot MABP - \beta \cdot \Phi(MABP) \cdot CBF(MABP), \qquad (6)$$

in which we have changed the overall strength of the feedback by multiplying the feedback function $\Phi$ by a constant denoted as $\beta$. In Eq. 5 $\beta$ was set as equal to 1.e. there is no change In Fig.3 we consider three cases: $\beta$=1, $\beta$=0.1, $\beta$=1.3. The $\beta$=0.1 case represents a strong suppression of the feedback;, as we shall see later, this is the typical situation in hypercapnia. The case of $\beta$=1.3 represents an increase in feedback to the point where CBF is suppressed and decreases instead of remaining relatively constant in the usual plateau region with increasing MABP. We found such behavior in some of Harper's (1966) dogs.). To our knowledge this effect has not been described in the literature. We suspect, extrapolating from hypercapnia (for which $\beta \ll 1$ ), through normocapnia (for which autoregulation is assumed, i.e. $\beta \sim 1$), that in hypocapnia $\beta > 1$.

RESULTS

**Analysis of Harper's (1966) data**

Harper (1966) measured CBF while changing MABP in a wide range 12 dogs, and published the experimental data for each of his experimental animals . To our knowledge, these are the only published data for individual animals in which CBF was measured over a wide range of MABP. As we shall see below, the parameters describing CBF in relation to MABP differed considerably among the animals, but for the sake of better statistics averaging the results may lead to incorrect conclusions.. For example Harper (1966) in his paper stated, "Over a fairly wide range of blood pressure (from 90 to 180 mm Hg) the blood flow remained relatively constant, despite a varying blood pressure. This phenomenon will hereafter be referred to as 'autoregulation'." When, however, we analyze each dog separately, we find rather large deviations.

When we analyze the data from the dogs on the basis of the simple model of autoregulation in Eq. 5 as modified by Eq. 6, we do not find the sharp raise in CBF at the higher levels of MABP. We, therefore, we apply the following model of the feedback function

$$\Phi(MABP) = \begin{cases} 0, & for \quad 0 \leq MABP \leq MABP_l \\ \beta \left( \dfrac{S \cdot MABP}{CBF(MABP_l)} - 1 \right), & for \quad MABP_l \leq MABP, \end{cases} \qquad (7)$$

where the constant β is introduced to indicate the change in feedback with respect to the ideal autoregulation, i.e. the case in which CBF does not change with MABP in the plateau region. For ideal autoregulation $\beta$=1. This model assumes that the feedback function remains linear, the differing from the ideal autoregulation only in the strength of the feedback.

This model has three parameters $S$, $MABP_1$, and $\beta$. $MABP_1$ is the threshold arterial mean blood pressure below which there is no feedback. Below $MABP_1$ CBF is dependent only on the slope parameter S, i.e., $CBF(MABP)=S \cdot MABP$. At the transition point $MABP_1$

$$CBF(MABP_1)= S \cdot MABP_1 . \qquad (8)$$

Substituting Eq. 8 into Eq. 7 we obtain

$$\Phi(MABP) = \begin{cases} 0, & for \quad 0 \leq MABP \leq MABP_1 \\ \beta \cdot \left( \dfrac{MABP}{MABP_1} - 1 \right) & for \quad MABP \geq MABP_1 \end{cases} \qquad (9)$$

while CBF is obtained by Eq.3. It should be noted that $\Phi(MABP)$ depends only on two parameters, $MABP_1$ and $\beta$. Moreover it is linear with respect to MABP, allowing a simple fit to the data.

The parameters were determined as follows. First, the parameter S was determined from the line starting from zero and tangential to the experimental points (i.e., not intersecting the lines connecting the experimental points). Next, the experimental values of $\Phi(MABP)$ were determined by Eq. 4 from the value of S and the experimental values of $CBF(MABP)$. The feedback function was obtained by a linear fit of Eq. 9 to the experimental data. . The parameters so determined individually for each of the 12 dogs are given in Table 1. Table 1 also includes the average values of $PaCO_2$ for each dog. In dogs B9-B12, which were hypercapnic , there was practically no autoregulation, and the CBF data could be fitted with strait lines (i.e., $\beta=0$).

One should note that in Table 1 the values of the parameter $\beta$ deviate considerably from the value $\beta=1$, which is characteristic of the condition of ideal (classical) autoregulation. It should also be noted that in two cases (Dogs B6 and B7) in which $\beta > 1$, CBF might decrease with increasing MABP in contrast with most other cases. .

For values of $PaCO_2$ in the range of 30.9-40.9 mm Hg there was for each individual dog no correlation between the values of the parameter $\beta$ and $PaCO_2$ (Table 1). Obviously, there should be a strong dependence of the parameter $\beta$ on $PaCO_2$ inasmuch as at high levels of $PaCO_2$, e.g., greater than 70 mm Hg, $\beta$ is practically zero. This indicates that the dependency of $\beta$ on $PaCO_2$ is different for each individual.

Except for Dogs B2 and B3, the results in all the other cases appears to support the hypothesis that the feedback function $\Phi(MABP)$ is linear with respect to MABP above a given threshold of MABP, i.e., $MABP_1$ . Inasmuch as there was no error analysis of Harper's experimental data, one cannot rule out that the experiments with Dogs B2-B3 had large errors.

## Other analyses

Harper (1966) presented the combined data obtained in 8 dogs, Dogs B1-B8. They are presented in Fig.7. The average of these data superficially seem to support the existence of classical autoregulation. We have indicated it by drawing the continuous line with a plateau through the very dispersed data. The great dispersion of the data is an indication that individual characteristics of each dog differ from the average of all . Indeed, Table 1 and Figs. 4 and 5 show this explicitly.

In Fig.8 we present the averaged data obtained from 5 Baboons by MacKenzie et al. (1979) Here again the impression is that on average the data seem to justify well classical autoregulation. Although this plot looks better than the one in Fig.7, one cannot rule out the possibility of large individual differences between the animals. Unfortunately, in the publication the data on the individual animals were not presented.  .

## DISCUSSION

In this paper we have analyzed the data obtained by Harper (1966) in 12 Dogs, in which  CBF was measured over a wide range of MABP. This is, to our knowledge, the only publication in which  the data for each individual animal were tabulated. Contrary to the belief that these data support the picture of classical autoregulation, i.e. that CBF is almost constant in the plateau region, we found a somewhat different picture. The analysis of the data for each animal separately indicates that large deviations from the classical autoregulation may exist. We were able to interpret these data by  a simple model   that is based on the following assumptions: 1)  up to a threshold level of MABP, denoted as $MABP_1$, CBF is directly proportional to MABP (as in a rigid pipe). Above $MABP_1$  up to a level of  $MABP_2$, at  which breakthrough occurs, there is a regulated suppression of CBF which can be explained by a negative feedback on CBF. This feedback is well described by a linear function of MABP (see Eq.  9) with a slope proportional to the parameter $\beta$ which may vary considerably among  different individuals. The classical autoregulation model with a plateau between $MABP_1$ and $MABP_2$  is a particular case of this model with $\beta=1$.

This  model describes quite well the results obtained  in dogs (Harper 1966) for which the individual feedback slope parameter varied to great extent, indicating the importance of using data obtained in individuals rather than the averaged data obtained for different individuals

Although autoregulation can be explained by the effect of a feedback suppressing the CBF, too much feedback may lead to such suppression that CBF, instead of increasing with increasing MABP, may, on the contrary, actually decrease. This is a new effect, not previously described , but it is predicted by our model and is actually seen in the data from Harper's Dogs B6 and B7 (Fig. 5).

The autoregulation, as well as the feedback, disappears when $PaCO_2$ is very highly elevated. This fact indicates that all the model parameters should be very sensitive to $PaCO_2$ although there are presently not enough data to describe this dependence. In any case, a good model of the regulation of CBF will depend not only on MABP, but also on $PaCO_2$, both of which have major influences on CBF. It is our intention to attempt to develop such a model.

# Acknowledgements

I am grateful to Dr. Louis Sokoloff for his interest, encouragement and many helpful comments essential for the completion of this work.

## Table 1

The parameters $S$, MABP$_1$, $\beta$ used for fitting the data in Fig. 4 for experiments done on Dogs B1-B12. The values of PaCO$_2$ were also added.

| Dog | $S$ (ml/min/100g/mmHg) | MABP$_1$ (mm Hg) | $\beta$ | av. PaCO$_2$ (mm Hg) |
|------|------|------|------|------|
| B1 | 0.0135 | 48 | 0.57 | 31.8 |
| B2 | 0.0082 | 75 | 0.49 | 31.9 |
| B3 | 0.0099 | 52 | 0.43 | 40.9 |
| B4 | 0.0128 | 27 | 0.35 | 37.9 |
| B5 | 0.0118 | 24 | 0.29 | 30.9 |
| B6 | 0.0139 | 68 | 1.18 | 34.3 |
| B7 | 0.0087 | 110 | 1.37 | 36.4 |
| B8 | 0.0188 | 41 | 0.72 | 38.2 |
| B9 | 0.0078 | -- | ~0 | 69.4 |
| B10 | 0.0116 | -- | ~0 | 74.9 |
| B11 | 0.0105 | -- | ~0 | 68.2 |
| B12 | 0.0094 | -- | ~0 | 86.1 |

# Figure captions

Figure 1. The classical picture of autoregulation. For MABP's in the range of

   60-130 mm Hg there is no change in the CBF.

Figure 2. The feedback function F of Eq. 5 as a function of MABP. In the plateau

   region it is linear.

Figure 3. The change in CBF as a result of the change in the feedback strength.

   The constant $\beta$ is a multiplication factor, which multiplies the feedback

   function F(MABP).

Figure 4. Experimental data of the feedback function $\Phi$ for each one of the B1-B8

   Dogs (Harper 1966). The fits are according to the parameters in Table 1.

Figure 5. Experimental data of CBF for each one of the B1-B8 Dogs (Harper 1966).

   The fits are according to the parameters of Table 1.

Figure 6. Experimental data of CBF for each one of the B9-B12 Dogs (Harper 1966).

   As practically the parameter $\beta$=0, only the linear fit to CBF is presented.

Figure 7. The experimental data (Harper 1966) of the 8 (B1-B8) Dogs taken

   together. On average the data seem to justify the classical autoregulation,

   depicted by the continuous line.

Figure 8. The experimental data of MacKenzie et al. (1979) on averaged CBF in 5

    baboons measured over a wide range of MABP. On

   average the data seem to support the classical autoregulation.

נבמחק ¶:

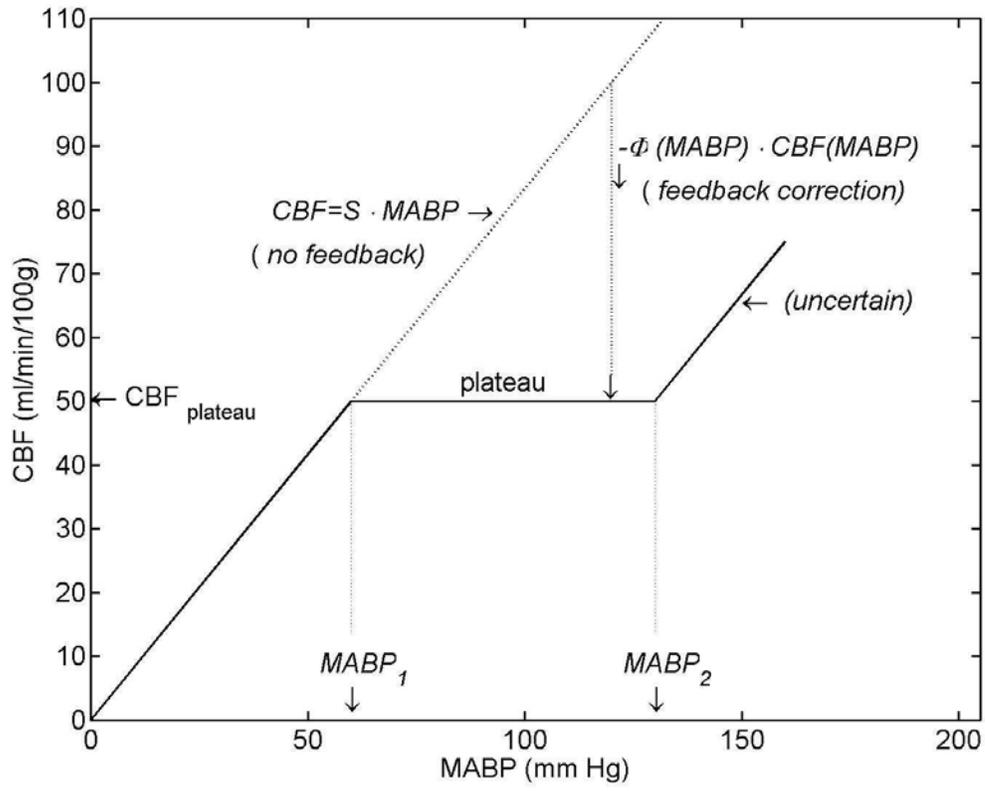

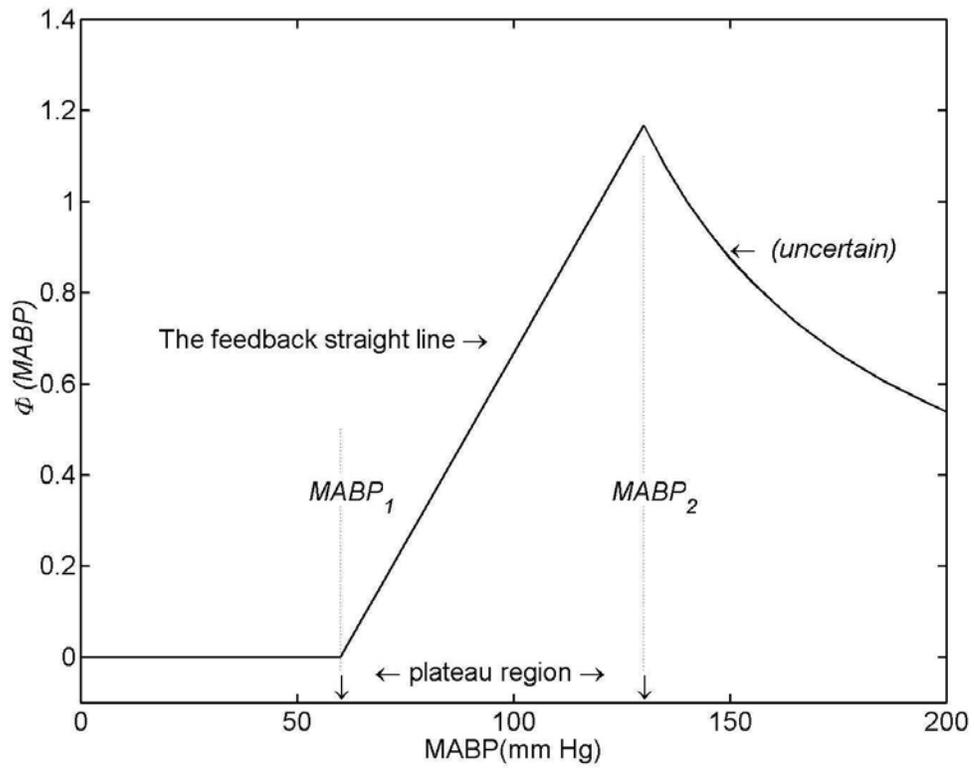

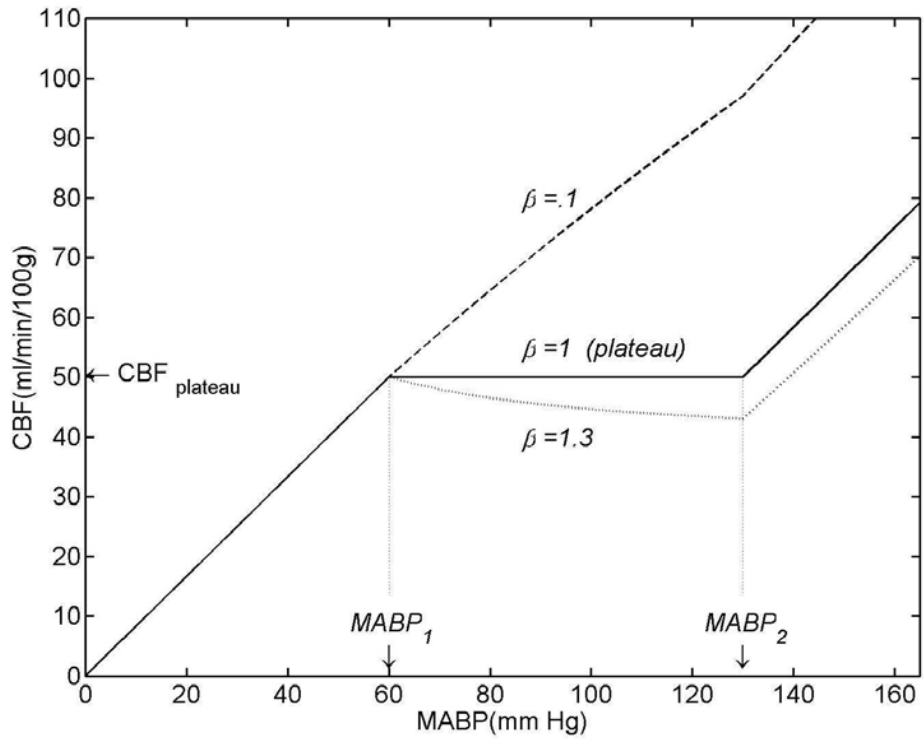

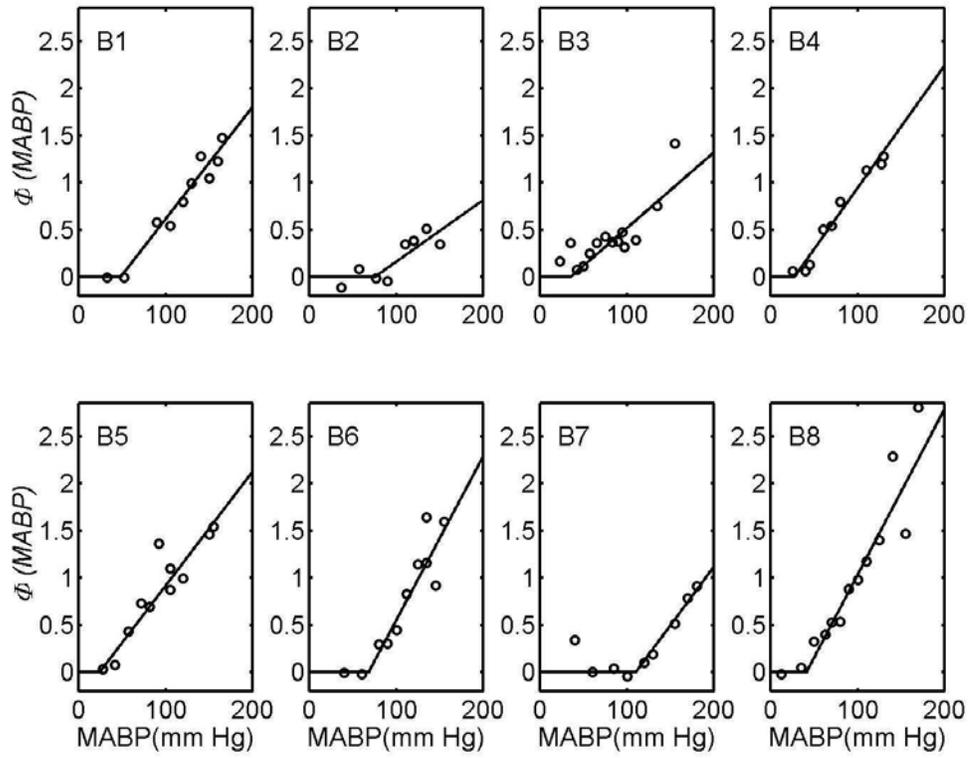

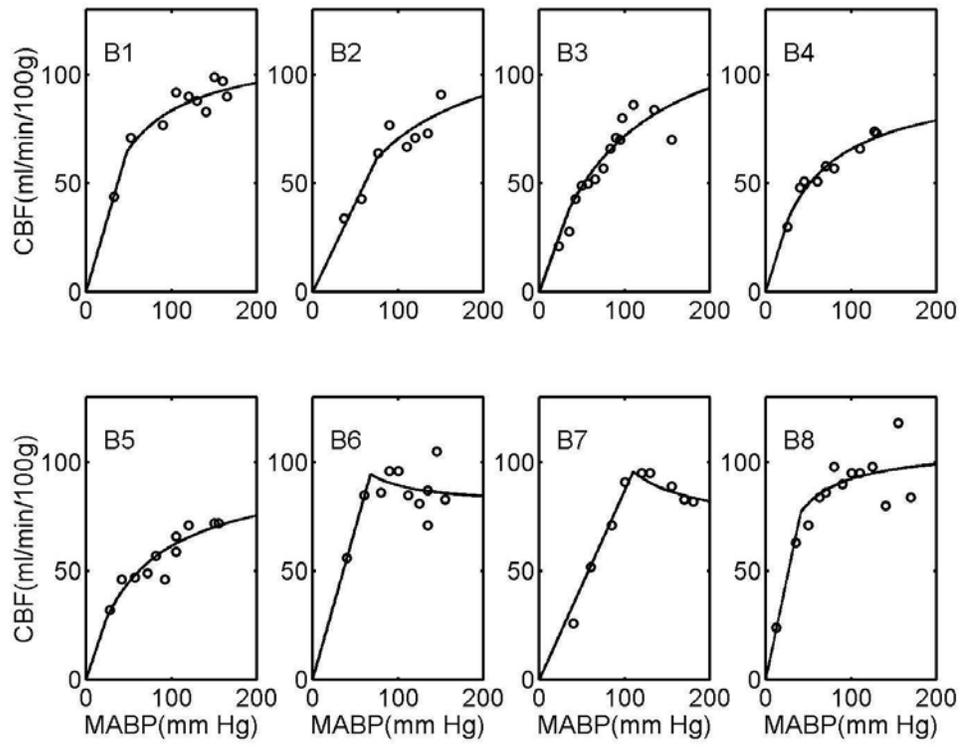

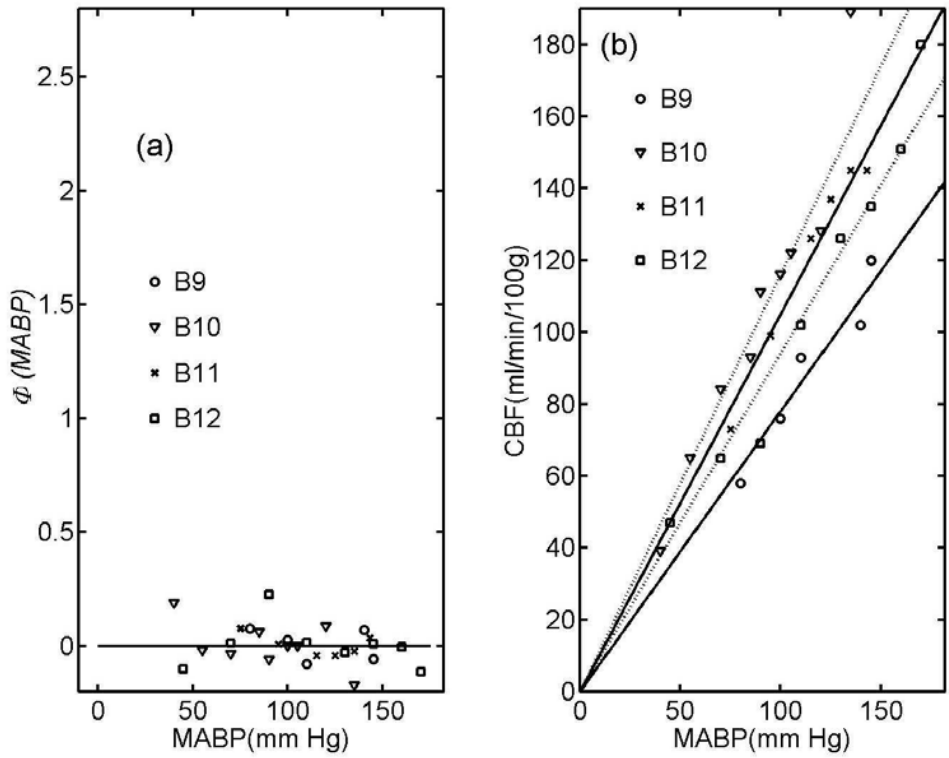

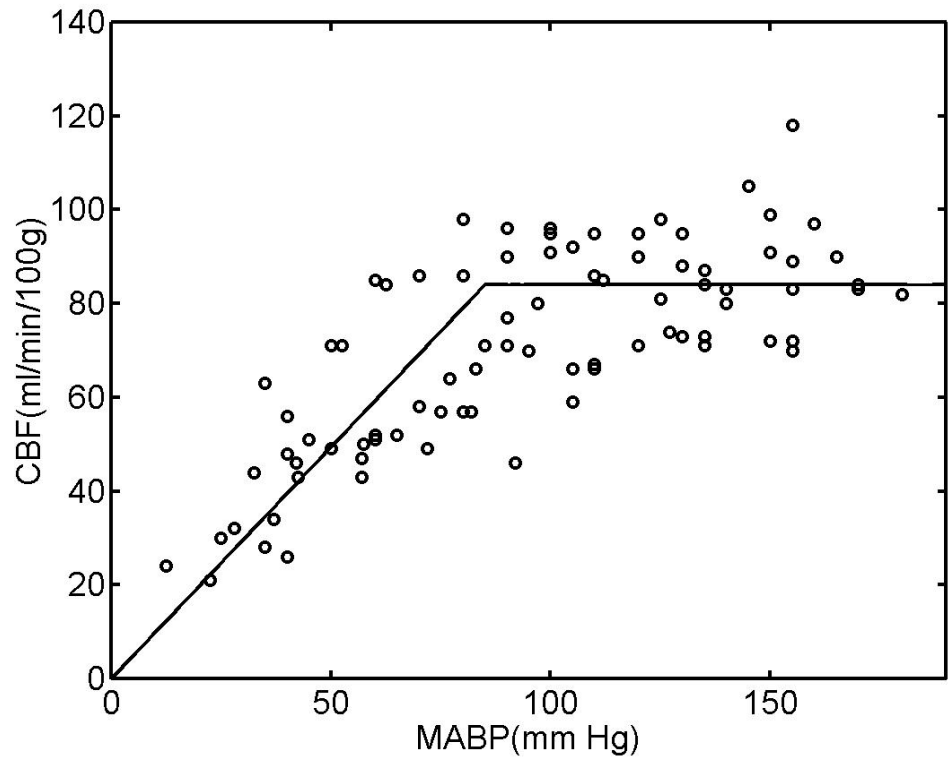

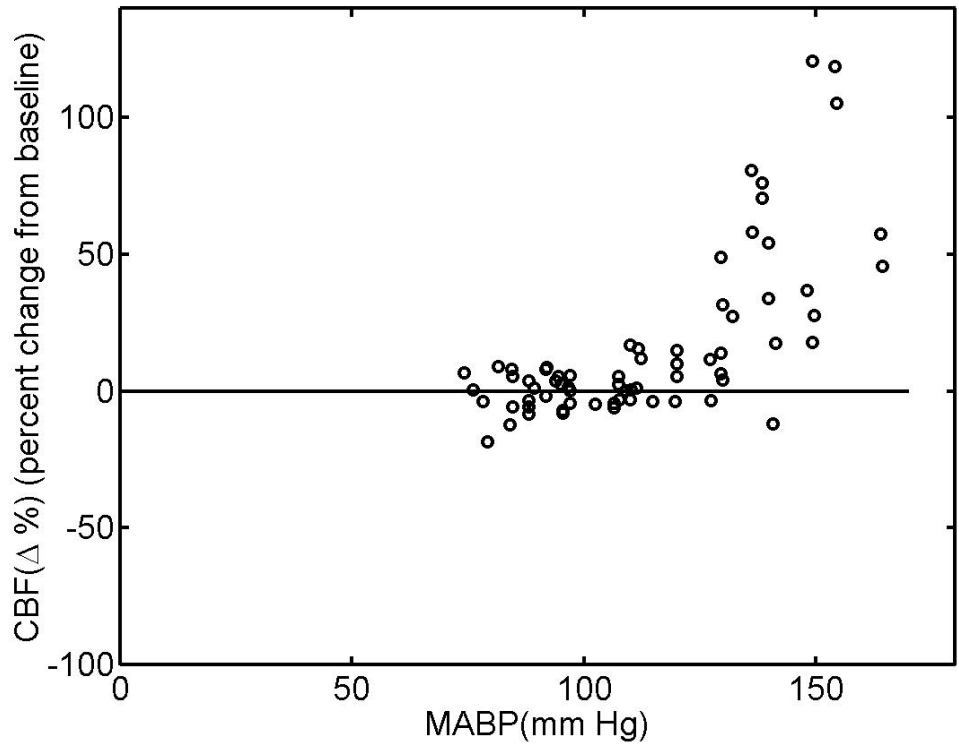